# Effects of Metallicity on Graphite, TiC, and SiC Condensation in Carbon Stars.

Gabrielle M. Adams[1,2], Katharina Lodders[2]

## Abstract

From transmission electron microscopy and other laboratory studies of presolar grains, the implicit condensation sequence of carbon-bearing condensates in circumstellar envelopes of carbon stars is (from first to last) TiC-graphite-SiC. We use thermochemical equilibrium condensation calculations and show that the condensation sequence of TiC, graphite, and SiC depends on metallicity in addition to C/O ratio and total pressure. Calculations were performed for a characteristic carbon star ratio of C/O = 1.2 from $10^{-10}$ to $10^{-4}$ bars total pressure and for uniform metallicity variations ranging from 0.01 to 100 times solar elemental abundances. TiC always condenses at higher temperatures than SiC, and the carbide condensation temperatures increase with both increasing metallicity and increasing total pressure. Graphite, however, can condense in a cooling circumstellar envelope before TiC, between TiC and SiC, or after SiC, depending on the carbon-bearing gas chemistry, which is dependent on metallicity and total pressure. Analytical expressions for the graphite, TiC, and SiC condensation temperatures as functions of metallicity and total pressure are presented. The inferred sequence from laboratory presolar grain studies, TiC-graphite-SiC, is favored under equilibrium conditions at solar and subsolar metallicities between ~$10^{-5}$ to $10^{-8}$ bar total pressure within circumstellar envelopes of carbon stars with nominal C/O = 1.2. We also explored the dependence of the sequence at C/O ratios of 1.1 and 3.0 and found that as the C/O ratio increases, the TiC-graphite-SiC condensation sequence region occurs towards higher total pressures and lower metallicities.

---

[1] Corresponding author a.gabrielle@wustl.edu
[2] Department of Earth, Environmental, & Planetary Sciences and McDonnell Center for the Space Sciences, Washington University in St. Louis, St. Louis, MO 63130, USA

1. Introduction

Presolar grains are nm-to-μm size grains that predate our solar system (Lodders & Amari 2005; Zinner 2014; Nittler & Ciesla 2016). Most of these grains were formed in stellar winds of asymptotic giant branch stars and supernova ejecta and avoided destruction during the gravitational collapse of the solar nebula. Carbonaceous dust such as graphite (C(Gr)), silicon carbide, (SiC), and refractory carbides such as titanium carbide (TiC) make up about 20 percent of identified presolar grains in the most primitive meteorites (Hoppe et al. 2022; Leitner et al. 2020). Many carbonaceous presolar grains condensed in the winds of carbon stars. These are asymptotic giant branch (AGB) stars of low mass (1.3-3.5 $M_\odot$) which have C/O ratios greater than unity and favor condensation of carbonaceous dust (Lodders & Fegley 1993, 1995, 1997a,b; Bernatowicz et al. 2006). Graphite grains are commonly found with TiC and/or other refractory carbide inclusions, often located in the center of the graphite, and these carbides likely served as nucleation centers (Bernatowicz et al. 1996; Croat et al. 2005; Amari et al. 2014). In contrast, SiC inclusions are rarely found in graphite grains; only a few instances are known (Croat et al. 2010; Bernatowicz et al. 1996). From these studies of presolar grains, the inferred condensation sequence for reduced carbonaceous condensates in carbon stars is (from first to last) TiC-C(Gr)-SiC.

Thermochemical equilibrium calculations can be used to narrow the temperature and pressure conditions that allow a particular condensation sequence for a given bulk elemental composition (e.g., for solar-like compositions: Grossman 1972; Grossman & Larimer 1974; Ebel & Grossman 2000; Lodders 2003; and for solar-like compositions with increased C/O: Lodders & Fegley 1993, 1995). Those results are for solar metallicity. However, condensation temperatures also depend on the overall elemental composition in the system, and changes from solar composition are described by stellar metallicity. Metallicity, often abbreviated as Z, is the mass fraction of all elements heavier than helium. For the Sun, Z is about 0.017, and the remainder of the mass fraction is He and H (Magg et al. 2022). Here we use the definition of metallicity often used in stellar spectroscopy, which is defined by the atomic ratio of a heavy element to hydrogen relative to the respective ratio in the Sun on a logarithmic scale[3]. This definition of metallicity can be directly related to the thermochemical relations describing the condensation temperatures (see Equation 24). The metallicity dependence of the TiC-C(Gr)-SiC sequence has not been investigated systematically beyond exploratory calculations reported in abstracts from Lodders & Fegley (1998) and Lodders (2006).

The carbon stars that provided presolar grains had to evolve to the AGB stage and release their dust into the interstellar medium before the solar system formed. Low-mass stars (~<4 $M_\odot$) that can become carbon stars require around 0.3 Ga (3-4 $M_\odot$) to about 5.5 Ga (1.3 $M_\odot$) to reach the AGB phase (Ekström et al. 2012), thus low mass stars formed in the same generation of stars as the Sun evolve too slow to provide grains to the solar system. Grain providing AGB stars must have formed at least ~0.3 - ~5.5 Ga before the Sun, and thus these stars plausibly came from generations of stars which were biased toward metal-poor (low-metallicity) stars. Though an age-metallicity relation is unreliable for individual stars, the general trend of metallicity increasing over time is seen in the stellar population of our galaxy (e.g., Lian et al. 2023). In our solar neighborhood, the metallicity density function (MDF) of stars peaks around solar metallicity regardless of stellar age, but older populations of stars have a much wider range in metallicity. Stars in the solar neighborhood that are older than the Sun (~5 Ga) have a subsolar metallicity tail in their MDF that extends to [M/H] = -1 (e.g., Casagrande et al. 2011; Bensby et al. 2014, Bergemann et al. 2014). While some fraction of stars in the MDF of presolar stars has supersolar metallicity, there is an additional complication for the case of supersolar metallicity carbon stars as presolar grain sources. In supersolar metallicity stars, the absolute amount of carbon needed to be produced and dredged up to reach the C-star phase is much greater than at subsolar and solar metallicities. The time required to produce and dredge up enough carbon to make a carbon star may then be longer than the AGB mass-loss phase, thus while the C/O ratio increases, it may not exceed C/O > 1, as is needed for carbonaceous grains to condense (Herwig 2005; Boyer et al. 2013; Straniero et al. 2023).

---

[3][M/H] ≡ log(N(M)/N(H))∗ − log(N(M)/N(H))☉, where N(M) is the number of atoms of M and N(M)/N(H) is the abundance of an element.

Gail et al. (2009) combined a chemical evolution of the Milky Way, a stellar evolution model, and a dust condensation model to set lower limits of [M/H] = -0.24 for SiC formation and [M/H] = -0.85 for graphite grain formation in presolar grain sources. Cristallo et al. (2020) used similar modelling and found that the majority of dust producing carbon stars had metallicities -0.2 ≤ [M/H] ≤ 0.

However, some studies have suggested supersolar metallicity carbon star contributions to presolar grains. Lewis et al. (2013) used Si-isotopic measurements of SiC grains in combination with galactic chemical evolution models and determined that the majority of SiC grain parent AGB stars had metallicities between 0 ≤ [M/H] ≤ +0.2. Lugaro et al. (2018, 2020) argued that large SiC grains came from AGB stars with metallicities as high as [M/H] = +0.3, based on nucleosynthesis modeling, as well as matching measured Sr, Zr, and Ba isotopic ratios in SiC grains to AGB star models and Ba star observations, assuming the Ba stars have inherited their composition from AGB star companions.

We investigate effect of metallicity on the TiC-C(Gr)-SiC condensation sequence to constrain the metallicities of the carbon stars where these grains originated and consider a wide range of metallicities to show trends in chemistry.

This paper is organized as follows: in §2 we describe the computations; in §3 we describe the results for the gas chemistry and condensation temperatures; In §4 we discuss our gas chemistry results in the context of other studies investigating the TiC-C(Gr)-SiC sequence, the implications of our results for parent star metallicities, and the astrophysical conditions of grain formation; and in §5 we summarize our conclusions.

## 2. Methods

The ideal gas chemical equilibrium calculations were performed using the CONDOR code (Lodders & Fegley 1993, 1995, 1997a,b; Fegley & Lodders 1994; Lodders 2003). The code simultaneously solves for mass balance, chemical equilibrium, and charge balance of all stable and long-lived naturally occurring elements (H to Bi, Th, U), as well as Pu and Tc, and their compounds between gases and condensed stable phases.

We use the solar photosphere abundances from Magg et al. (2022) where available and otherwise from Palme et al. (2014) and references therein. The observed in C-stars typically range from 1 to 1.3 (e.g., Lambert et al. 1986, Abia et al. 2002, Abia et al. 2024). The C/O ratio adopted for C-type stars in our nominal case was 1.2, calculated by enhancing the adopted solar carbon abundance. We also investigate C/O ratios of 1.1 and 3.0 in §3.2.4. Total pressure was varied from $10^{-4}$ to $10^{-10}$ bar, which covers the ranges reported in observational studies by Wong et al. (2016), Khouri et al. (2016, 2018, 2024), Vlemmings et al. (2017, 2019), Ohnaka et al. (2017), and Adam & Ohnaka (2019). Metallicity distributions in the solar neighborhood range from 0.1 to 3 times solar (e.g., Casagrande et al. 2011; Bensby et al. 2014, Bergemann et al. 2014). We vary metallicity from 0.01 to 100 times solar to investigate the trend in condensation sequence over a wide range of parameters.

The major data sources are the 2nd to 4th editions of the NIST-JANAF Thermochemical Tables (Stull & Prophet 1971; Chase 1985; Chase 1998), and the IVTAN Tables (Belov et al. 1999). Other data sources are listed within Lodders & Fegley (1993, 1995) and Fegley & Lodders (1994). Data for several compounds that were erroneously transcribed from the 2nd to the 3rd and 4th editions of the JANAF tables or had computational errors in the basic thermochemical properties are corrected in Lodders (1999, 2004). Gundiah et al. (1996) identified and corrected the standard Gibbs energy of formation, $\Delta_f G°$, and equilibrium constant, K, for $C_2H_2$(g) that were erroneously transcribed from the 2nd to the 3rd edition of JANAF and subsequently corrected in the 4th edition. Particularly important for this work are the thermochemical properties for $C_2H$(g). The 2nd, 3rd, and 4th editions of JANAF (Stull & Prophet 1971; Chase 1985; Chase 1998) give the standard heat of formation as $\Delta_f H° =$ +476.976 kJ/mol based only on experimental measurements done prior to 1967. Taking more recent experimental measurements into account, Dorofeeva & Gurvich (1992) and Gurvich et al. (1993) give +569 kJ/mol, making $C_2H(g)$ significantly less stable. This difference is important as several prior condensation calculations reported $C_2H$ as an important gas (see §4.1).

In the following, we illustrate the basic gas and solid reaction equilibria for graphite condensation. Though thermochemical equilibrium is path independent, it is useful to consider specific major reactions that help to explain how changes in metallicity, total pressure, and C/O ratio affect the gas chemistry and condensation temperatures. These exemplary reactions can often yield good approximations, however, all results reported are for the full set of equilibria, including thermal ionization, for all elements and compounds in the code.

The total number of carbon atoms in all gases containing C is $N(C_{Tot})$:

$$N(C_{Tot}) = N(CO) + N(C) + N(CH) + 2N(C_2H_2) + 3N(C_3) + \ldots$$

(1)

Note that a stoichiometric factor "η" must appear for molecules containing η atoms of carbon in this sum of all carbon atoms. Neglecting removal of carbon from the gas into condensates for now, dividing Equation (1) by the sum of all gas molecules gives the ratio of the total number of carbon atoms to the total number of gas species:

$$C_{\Sigma X} = \frac{N(C_{Tot})}{N(H) + N(H_2) + N(He) + N(Ne) + N(N_2) + N(CO) + N(C) + N(C_2H_2) + N(C_3) + \ldots}$$

(2)

$$\approx \frac{N(C_{Tot})}{N(H) + N(H_2) + N(He)}$$

(3)

The sum "$N(H) + N(H_2) + \ldots$" is the sum of all gas species from all elements (ions, monatomics, and molecules). Hydrogen and helium are the two major elements with elemental abundances much greater than those of any other element at solar and subsolar metallicities, so the sum in the denominator can be approximated as "$N(H) + N(H_2) + N(He)$". The N(H) and N(H$_2$) are obtained from the temperature dependent 2H(g)= H$_2$(g) equilibrium and the adopted mass balance approximation that N(H$_{Tot}$) = N(H) + 2N(H$_2$). This is a good approximation because thermal ionization of monatomic H is insignificant relative to H and H$_2$ abundances at the pressure and temperatures considered here.

With Equation (1), we then have for the sum of mole fractions of all carbon-bearing gases:

$$C_{\Sigma X} = X_{CO} + X_C + 2X_{C_2H_2} + 3X_{C_3}...$$

(4)

Note that the η (= 2 and = 3 in the example here) ensures mass balance. Multiplying the mole fractions by the total pressure, P$_{Tot}$, gives the partial pressures for each C-bearing gas:

$$C_{\Sigma P} = C_{\Sigma X} P_{Tot} = P_{Tot}(X_{CO} + X_C + 2X_{C_2H_2} + 3X_{C_3}...) = P_{CO} + P_C + 2P_{C_2H_2} + 3P_{C_3}...$$

(5)

Each partial pressure can be expressed by the corresponding equilibrium constant expression, e.g.

$$P_{C_2H_2} = X_{C_2H_2} P_{Tot} = K_{C_2H_2} fH_2 a_{Gr}^2$$

(6)

Where a$_{Gr}$ is the activity of graphite, the reference state for carbon. Activity is the effective concentration of a chemical in solution. For pure condensed phases in their standard state, the activity is unity. The equilibrium constant, K$_i$, is for the formation of molecule *i* from the constituent elements in their reference states with respect to standard state pressure (1 bar) and the conventional reference temperature (298.15 K). The values for K$_i$ are tabulated in the thermochemical data sources described previously. For gases, the reference variable is the gas fugacity, denoted by f.

The equation for the gas phase abundance of carbon from all gases containing carbon is:

$$C_{\Sigma P} = C_{\Sigma X} P_{Tot} = a_{Gr}[K_{CO} fO_2^{0.5} + K_C + 2K_{C_2H_2} fH_2 a_{Gr} + 3K_{C_3} a_{Gr}^2 +...]$$

(7)

One such equation must be written for each element, where the base variable is the activity or fugacity of each element in its reference state. These mass balance equations are nonlinear and coupled with each other through compounds that contain elements in common. The total mass balance equation for carbon in CONODR considers 276 C-bearing gases.

Graphite condensation occurs when the activity, $a_{Gr}$, is unity. The mass balance equation is then adjusted by a factor (1-α), where α is the fraction of total carbon condensed, using the criterion that the activity of graphite cannot exceed unity. Removal of carbon, Ti, and/or Si from the gas by TiC and/or SiC condensation is accounted for in a similar manner.

The CONDOR code solves mass balance equations iteratively until convergence. The convergence criteria requires that the abundance of each element in the gas phase plus all elements in condensates agree within 1 part in $10^9$ or less to the input abundances to account for mass balance of all species for each independently computed P, T, and metallicity point.

3. Results

We first describe results for hydrogen and carbon gas chemistry which are important for understanding the condensation sequence dependencies on metallicity and total pressure. Then, the condensation temperature curves for graphite, TiC, and SiC are reported. All temperatures reported are initial condensation temperatures. The graphite condensation temperatures are examined in detail and the importance of the carbon gas chemistry on the condensation sequence is shown. Finally, the metallicity and total pressure dependence of graphite, TiC, and SiC condensation is investigated at different C/O ratios.

3.1 Gas Chemistry

Gas chemistry refers to the gas phase composition resulting from the net thermochemical reactions, which change because of changing temperature, metallicity, total pressure, or C/O ratio. The hydrogen chemistry is relatively simple in the temperature, pressure, and metallicity ranges where graphite condensation occurs. Figure 1 shows the hydrogen gas speciation and the mole fraction of CO, which is always the most abundant C-bearing gas. The CO mole fraction increases by approximately a factor of two as the reaction $2H(g) = H_2(g)$ occurs with decreasing temperature because the denominator term in Equation (2) becomes smaller as H atoms recombine to $H_2$. At both pressures shown, $\log(P/\text{bar}) = -4$ and $\log(P/\text{bar}) = -10$, increasing metallicity increases the CO mole fraction because $N(C_{Tot})$ and $N(O_{Tot})$ are increased.

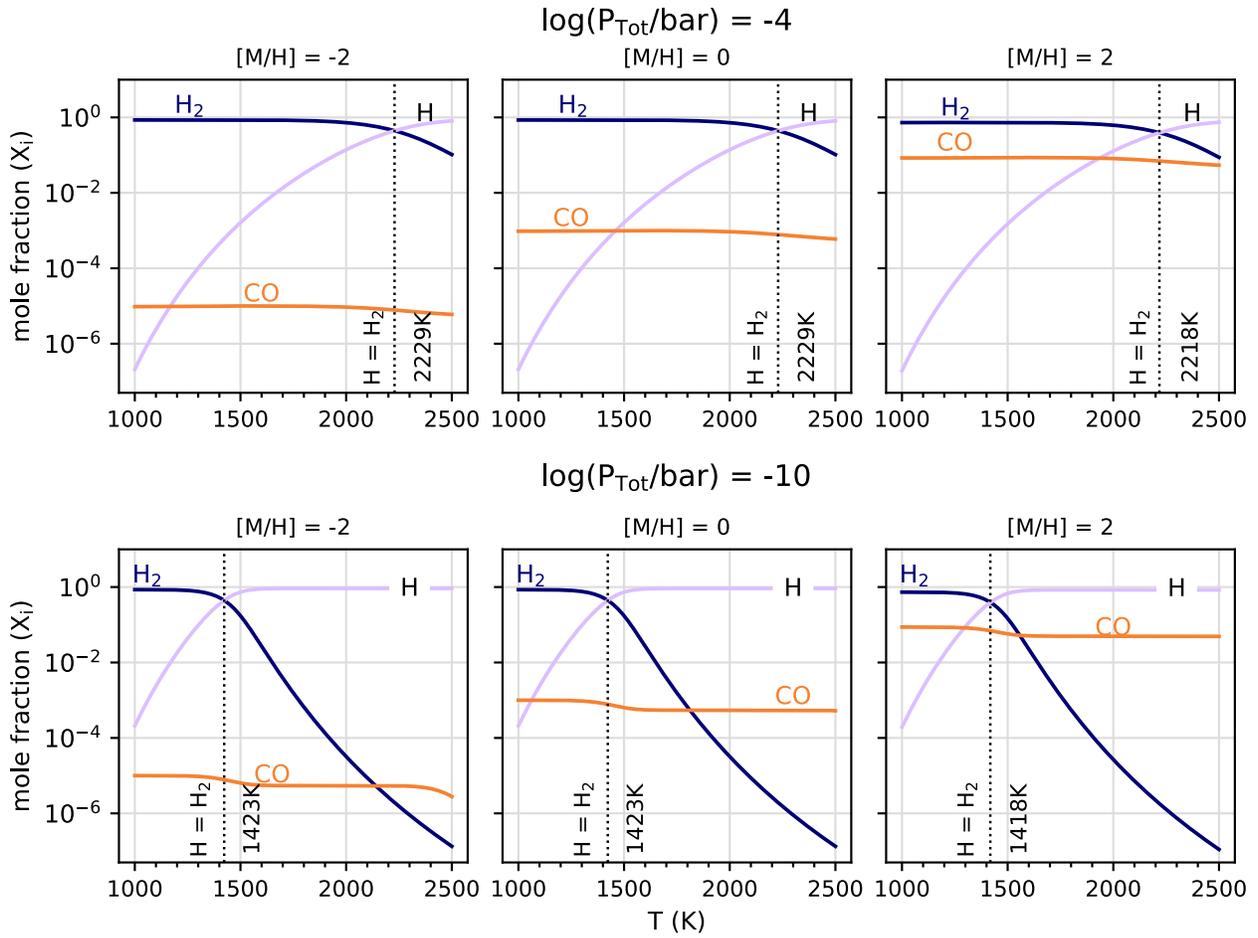

Figure 1. Gas chemistry of major species for C/O = 1.2 and select metallicities at log(P/bar) = -4 (top row) and log(P/bar) = -10 (bottom row). The temperature where $X(H) = X(H_2)$ is marked with a vertical dotted line. Monatomic H is more abundant than $H_2$ at higher temperatures.

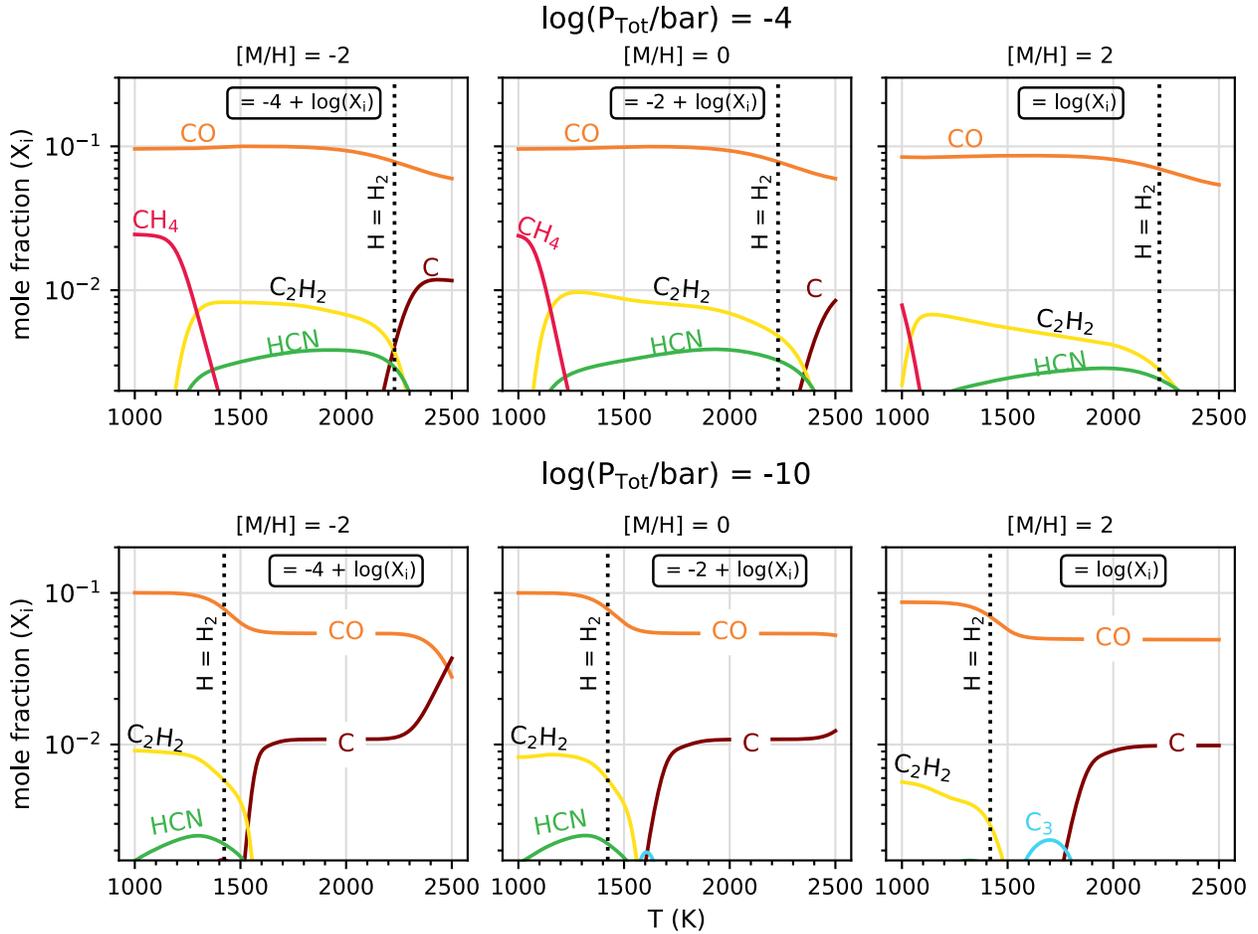

*Figure 2. Major carbon-bearing gases for C/O = 1.2 and select metallicities at log(P/bar) = -4 (top row) and log(P/bar) = -10 (bottom row). A scale factor for the mole fractions is indicated in each plot to allow comparison between speciation. The dashed black line is for X(H)= X($H_2$), as in Figure 1.*

In Figure 2 we show the other major C-bearing gases in addition to CO. Each metallicity panel is scaled by the order of magnitude indicated in the label to allow for comparison between metallicities. The gases shown are C, $CH_4$, $C_2H_2$, and HCN; other C-bearing gases ($C_3$, CS, CN) are usually less abundant by more than a factor of 100. At supersolar metallicities and low pressures, $C_3$ becomes increasingly abundant with increasing metallicity. The dotted line indicates where $H_2$ and H gases are equal in abundance. The change from H to $H_2$ as the major hydrogen gas with decreasing temperature occurs close, but not exactly at the temperature where $C_2H_2$ replaces monatomic C as the second major C-bearing gas.

Because CO has the highest bond energy of any known molecule, CO is more stable than any other C-bearing gas. Hence the second most abundant C-bearing gas is important for graphite condensation. In the higher pressure (log(P/bar) = -4) case (Figure 2), the relative $CH_4$ and C abundances are lower with increasing metallicity because increasing metallicity increases the number of carbon atoms available to form the carbon rich species $C_2H_2$ and nitrogen atoms available to form HCN (note that the total hydrogen abundance is metallicity independent). As metallicity increases, $C_2H_2$ is favored over other carbon rich molecules like $C_2$ or $C_2H$ because of its higher stability. This increase in carbon and nitrogen rich molecules is the result of Le Chatelier's principle driving gas chemistry toward larger molecules as the elemental abundances of C and N are increased (at constant P, T). In the bottom panel (log(P/bar) = -10), the highest metallicity (and therefore most carbon rich) shows a $C_3$ region that forms in the monatomic H dominated region.

Over the P-T and metallicity ranges shown in Figure 2, which are the same ranges used for the full set of calculations, the most important C-bearing gases at temperatures where graphite condensation begins are monatomic C and $C_2H_2$, and to a lesser extent, $C_3$. In the total pressure range considered here, methane ($CH_4$) is never the second most abundant C-bearing gas at the temperatures where graphite begins condensing. In the next section we show that the TiC-C(Gr)-SiC condensation sequence depends on the carbon and hydrogen gas chemistry described above.

3.2 Condensation Temperatures

The initial condensation temperatures, $T_{cond}$, for TiC, SiC, and graphite at C/O = 1.2 and pressures of $-10 \leq \log(P/bar) \leq -4$ are plotted in Figure 3 and listed in Table 1. TiC and SiC condensation temperatures increase with increasing total pressure and metallicity. Carbide condensation temperatures follow approximately the same curvature, and TiC always condenses at higher temperatures than SiC. Unlike TiC and SiC, graphite condensation can be pressure independent, decrease with increasing pressure, or increase with increasing pressure. Where graphite condensation temperatures increase with increasing pressure, the curvature is approximately the same as the curvature of the carbide condensation temperature curves, thus the sequence will always be C(Gr)-TiC-SiC. There is only a small pressure range for a given metallicity where graphite condensation temperatures decrease with increasing pressure. This pressure range is too small for intersection with the carbide condensation temperature curves to occur, and the sequence remains C(Gr)-TiC-SiC. Where graphite condensation temperatures are pressure independent, carbide condensation temperatures continue to increase with increasing pressure. At sufficiently high pressures, the carbide condensation temperatures can be higher than graphite condensation temperatures, changing the sequence. Generally, lower metallicity favors the pressure independent case for graphite condensation temperatures, thus favoring TiC-C(Gr)-SiC and TiC-SiC-C(Gr) sequences. The gas chemistry changes that cause the different pressure dependencies for graphite condensation are examined in §3.2.1.

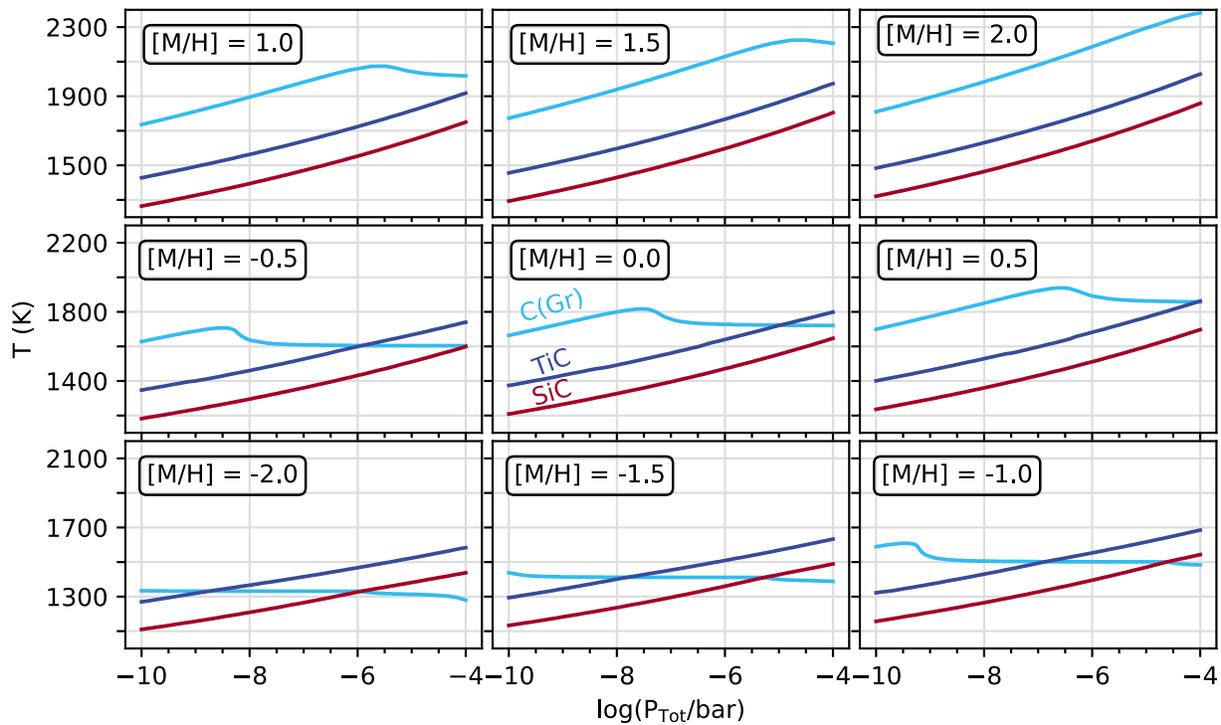

Figure 3. TiC, SiC, and graphite initial condensation temperatures at selected metallicities over a total pressure range -10 ≤ log(P/bar) ≤ -4, for C/O = 1.2.

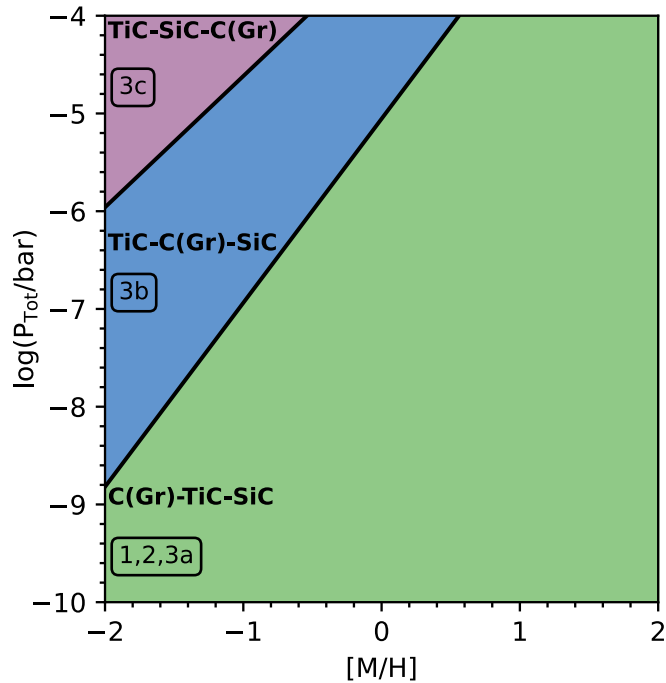

*Figure 4. Graphite, TiC, and SiC condensation sequence map for a C/O ratio of 1.2. The sequence boundaries are the intersection points of the TiC-graphite and SiC-graphite curves, as in Figures 3 and 5. The numbers in boxes indicate which region of Figure 5 corresponds to the condensation sequence. See §3.2.4 and §3.3 to estimate this plot at other C/O ratios.*

Figure 4 shows the condensation sequences that are favored at a given total pressure and metallicity. The condensation sequence predicted from presolar grains, TiC-C(Gr)-SiC, is possible at intermediate to high pressures and subsolar to slightly supersolar metallicity. Both total pressure and metallicity must be known to predict the sequence. §3.2.1 and Figure 5 assign five regions (R1, R2, R3a, R3b, R3c) to explain the changing condensation sequence. These five regions are indicated within the sequence they correspond to on Figure 4 for reference between Figures 4 and 5.

The boundaries in Figure 4 correspond to total pressures and metallicities where condensation curves (shown in Figure 5) intersect, and condensation temperatures are equal for TiC and graphite, approximated as

$$\log(P_{Tot}) = 1.8854\,[M/H] - 4.9909$$

(8)

And for equal graphite and SiC condensation temperatures in Figure 4 we approximate the boundary as:

$$\log(P_{Tot}) = 1.3370\,[M/H] - 3.2839$$

(9)

These approximations give the total pressure where the intersection of these points occurs within $\Delta\log(P/bar) = \pm\,0.05$ for a given metallicity.

*Table 1. Initial condensation temperatures in K for –2.0 ≤ [M/H] ≤ +2.0 and –10 ≤ log(P/bar) ≤ –4.*

| log(P/bar) | | | | | [M/H] | | | | |
|---|---|---|---|---|---|---|---|---|---|
| Graphite | -2.0 | -1.5 | -1.0 | -0.5 | 0.0 | 0.5 | 1.0 | 1.5 | 2.0 |
| -4.0 | 1280 | 1388 | 1484 | 1604 | 1721 | 1858 | 2018 | 2207 | 2382 |
| -4.5 | 1307 | 1393 | 1494 | 1604 | 1722 | 1860 | 2025 | 2224 | 2346 |
| -5.0 | 1314 | 1400 | 1501 | 1604 | 1723 | 1863 | 2042 | 2214 | 2295 |
| -5.5 | 1318 | 1411 | 1501 | 1604 | 1725 | 1870 | 2074 | 2176 | 2241 |
| -6.0 | 1331 | 1411 | 1501 | 1605 | 1728 | 1892 | 2061 | 2129 | 2186 |
| -6.5 | 1331 | 1411 | 1502 | 1607 | 1734 | 1938 | 2024 | 2081 | 2133 |
| -7.0 | 1331 | 1411 | 1503 | 1609 | 1757 | 1923 | 1982 | 2032 | 2081 |
| -7.5 | 1331 | 1411 | 1504 | 1615 | 1817 | 1889 | 1938 | 1985 | 2031 |
| -8.0 | 1331 | 1412 | 1505 | 1637 | 1800 | 1850 | 1895 | 1939 | 1983 |
| -8.5 | 1331 | 1413 | 1510 | 1707 | 1768 | 1811 | 1853 | 1895 | 1937 |
| -9.0 | 1332 | 1414 | 1531 | 1689 | 1733 | 1773 | 1813 | 1853 | 1893 |
| -9.5 | 1333 | 1419 | 1608 | 1660 | 1698 | 1735 | 1773 | 1812 | 1851 |
| -10.0 | 1334 | 1438 | 1589 | 1628 | 1664 | 1699 | 1736 | 1773 | 1810 |
| TiC | -2.0 | -1.5 | -1.0 | -0.5 | 0.0 | 0.5 | 1.0 | 1.5 | 2.0 |
| -4.0 | 1583 | 1633 | 1685 | 1740 | 1799 | 1862 | 1918 | 1973 | 2028 |
| -4.5 | 1553 | 1600 | 1650 | 1703 | 1759 | 1815 | 1866 | 1918 | 1969 |
| -5.0 | 1523 | 1569 | 1616 | 1667 | 1721 | 1768 | 1817 | 1865 | 1913 |
| -5.5 | 1495 | 1538 | 1584 | 1633 | 1680 | 1724 | 1769 | 1815 | 1860 |
| -6.0 | 1467 | 1509 | 1553 | 1599 | 1640 | 1682 | 1724 | 1767 | 1809 |
| -6.5 | 1441 | 1481 | 1523 | 1563 | 1598 | 1636 | 1681 | 1721 | 1761 |
| -7.0 | 1415 | 1454 | 1493 | 1527 | 1561 | 1597 | 1640 | 1678 | 1716 |
| -7.5 | 1390 | 1428 | 1461 | 1493 | 1526 | 1561 | 1601 | 1637 | 1672 |
| -8.0 | 1366 | 1401 | 1430 | 1460 | 1492 | 1529 | 1563 | 1597 | 1631 |
| -8.5 | 1343 | 1372 | 1400 | 1429 | 1464 | 1495 | 1527 | 1560 | 1592 |
| -9.0 | 1319 | 1345 | 1372 | 1400 | 1433 | 1462 | 1493 | 1524 | 1554 |
| -9.5 | 1294 | 1319 | 1345 | 1375 | 1403 | 1431 | 1460 | 1489 | 1518 |
| -10.0 | 1270 | 1294 | 1322 | 1347 | 1374 | 1401 | 1428 | 1456 | 1484 |
| SiC | -2.0 | -1.5 | -1.0 | -0.5 | 0.0 | 0.5 | 1.0 | 1.5 | 2.0 |
| -4.0 | 1438 | 1489 | 1543 | 1599 | 1647 | 1697 | 1750 | 1805 | 1859 |
| -4.5 | 1410 | 1458 | 1509 | 1554 | 1599 | 1647 | 1697 | 1749 | 1800 |
| -5.0 | 1382 | 1427 | 1470 | 1511 | 1554 | 1599 | 1646 | 1695 | 1744 |
| -5.5 | 1354 | 1395 | 1432 | 1470 | 1511 | 1554 | 1599 | 1645 | 1691 |
| -6.0 | 1327 | 1360 | 1395 | 1432 | 1471 | 1511 | 1553 | 1597 | 1640 |
| -6.5 | 1295 | 1327 | 1360 | 1395 | 1432 | 1470 | 1511 | 1552 | 1593 |
| -7.0 | 1265 | 1295 | 1327 | 1360 | 1395 | 1432 | 1470 | 1509 | 1548 |
| -7.5 | 1236 | 1265 | 1295 | 1327 | 1360 | 1395 | 1431 | 1468 | 1505 |
| -8.0 | 1209 | 1236 | 1265 | 1295 | 1327 | 1360 | 1394 | 1430 | 1464 |
| -8.5 | 1183 | 1209 | 1236 | 1265 | 1295 | 1327 | 1359 | 1393 | 1426 |
| -9.0 | 1157 | 1183 | 1209 | 1236 | 1265 | 1295 | 1326 | 1358 | 1389 |
| -9.5 | 1133 | 1157 | 1183 | 1209 | 1236 | 1265 | 1294 | 1325 | 1354 |
| -10.0 | 1110 | 1133 | 1157 | 1182 | 1209 | 1236 | 1264 | 1293 | 1321 |

### 3.2.1 Graphite Condensation

The shape of the graphite initial condensation temperature, $T_{cond}$ C(Gr), curve defines three distinct regions (R1, R2, and R3) as shown in Figure 5A-C; R3(a-c) is subdivided based on the condensation sequence. The same condensation curves are plotted in Figure 6 with the carbon and hydrogen gas speciation boundaries. The pressure dependence of the $T_{cond}$ C(Gr) curve is mainly controlled by the second most abundant C-bearing gas and most abundant hydrogen gas. The boundaries of the highlighted R2 are where changes in gas chemistry occur. There is no gas chemistry change in R3; the subdivisions are defined by where condensation curves intersect. These intersection points define the boundaries marked in Figure 4.

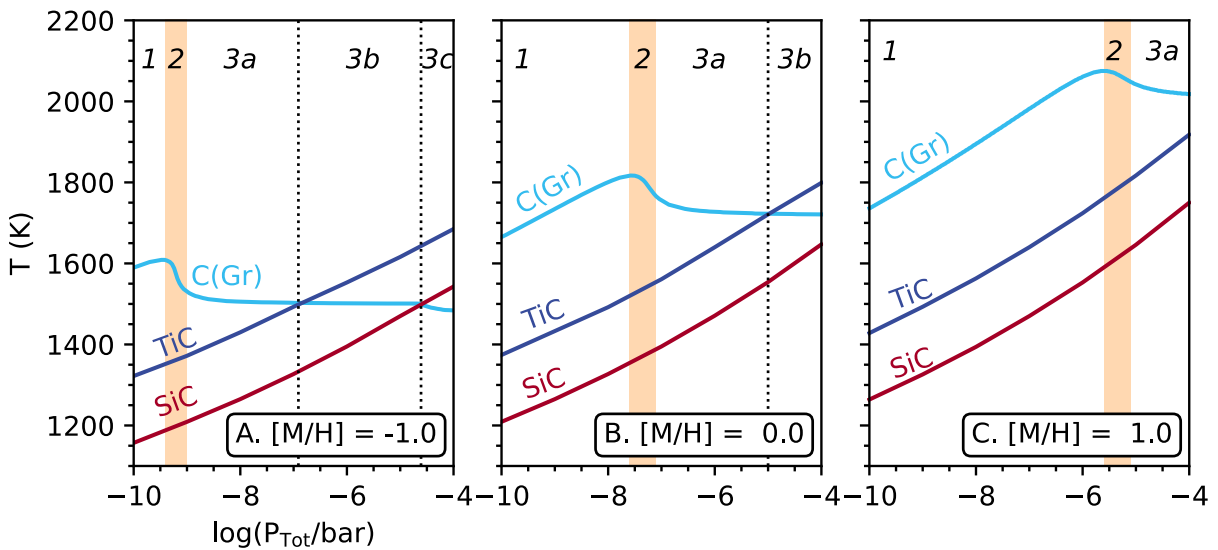

*Figure 5A-C. TiC, SiC, and graphite initial condensation temperatures at select metallicities over a total pressure range -10 ≤ log(P/bar) ≤ -4, for C/O = 1.2. The bounds of the shaded R2 region indicate where gas chemistry changes occur (see text). Subregions R3a, R3b, and R3c indicate where the condensation sequence is C(Gr)-TiC-SiC, TiC-C(Gr)-SiC, and TiC-SiC-C(Gr)*

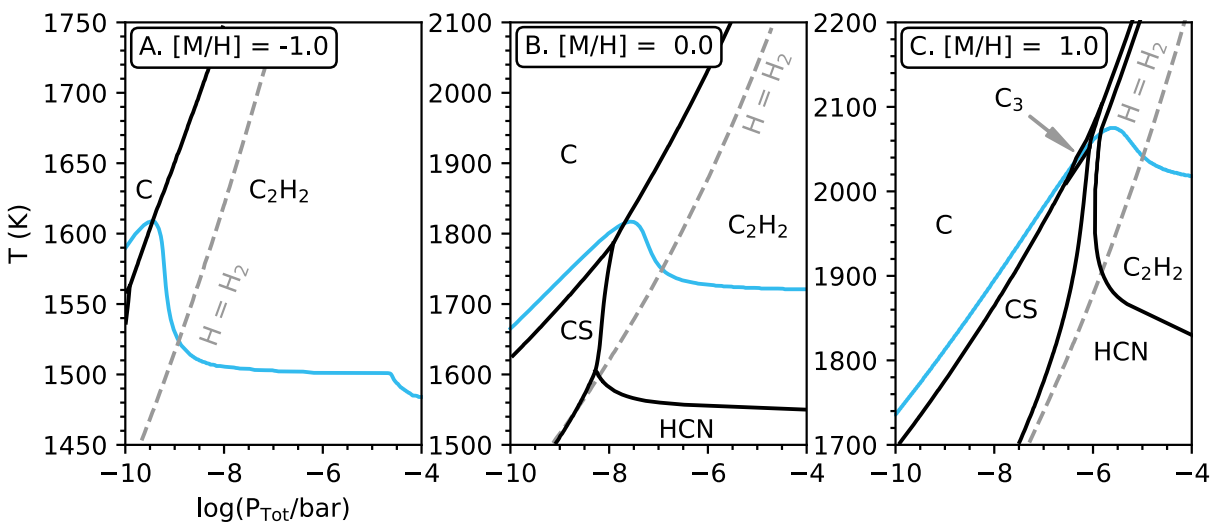

*Figure 6A-C. Graphite initial condensation temperature curves (blue curves) overlaid with boundaries showing the most abundant carbon-bearing gas species after CO, and the X(H) = X(H$_2$) boundary (dashed curves). The black curves are where mole fractions of two carbon-bearing gases are equal.*

In R1, the second most abundant C-bearing gas species is C(g) and graphite condenses via the net thermochemical reaction:

$$C(g) = C(Gr)$$

An approximate analytical solution for the graphite initial condensation temperature where monatomic C is the second most abundant C-bearing gas is derived below.

One approximation mentioned in §2 is that the denominator in Equation (2) is taken as the sum N(H$_{Tot}$) + N(He). Where H$_2$ is the dominant hydrogen species, N(H$_2$) = 0.5×N(H$_{Tot}$). Monoatomic He is always the most abundant He gas, so N(He) can always be used in the mole fraction denominator. The definition of the mole fraction of an element (E) depends on the hydrogen speciation and endmembers are:

$$X(E) = \frac{N(E)}{N(H_{Tot}) + N(He)}$$

(10)

$$\text{or } X(E) = \frac{N(E)}{N(H_2)+N(He)} = \frac{N(E)}{(0.5 \times N(H_{Tot}))+N(He)}$$

(11)

To a good first order approximation, all oxygen is in CO gas. The abundance of "available" carbon that can condense into graphite is effectively:

$$N(C) - N(O) = N'(C)$$

(12)

Graphite condensation occurs when the activity of graphite is unity. The expression for $K_{Gr}$, the equilibrium constant for formation of graphite from monatomic C(g) is:

$$K_{Gr} = \frac{a_{Gr}}{P_C}$$

(13)

Note that $K_{Gr}$ is the inverse of the equilibrium constant for formation of monatomic C gas from the graphite reference state.

In R1, the abundance of C(g) is approximately equal to that of all available carbon (i.e., C not in CO). This approximation holds best at pressures far from the equal abundance lines in Figure 6, but at worst it is off by a factor of two on the equal abundance lines.

In R1, the partial pressure, $P_C$, is equal to the mole fraction of available C multiplied by total pressure,

$$K_{Gr} = \frac{a_{Gr}}{X'(C) \times P_{Tot}}$$

(14)

where X'(C) is the molecular fraction of available carbon,

$$X'(C) = \frac{N'(C)}{N(H_{Tot})+N(He)}$$

(15)

Substituting Equation (15) into Equation (14):

$$K_{Gr} = \frac{a_{Gr}}{P_{Tot}} \times \frac{N(H_{Tot}) + N(He)}{N'(C)}$$

(16)

Expanding by $N(H_{Tot})/N(H_{Tot})$,

$$K_{Gr} = \frac{a_{Gr}}{P_{Tot}} \times \frac{N(H_{Tot}) + N(He)}{N'(C)} \times \frac{N(H_{Tot})}{N(H_{Tot})}$$

(17)

$$K_{Gr} = \frac{a_{Gr}}{P_{Tot}} \times \frac{1 + \frac{N(He)}{N(H_{Tot})}}{\frac{N'(C)}{N(H_{Tot})}}$$

(18)

Expressing the atomic abundances for any element E relative to hydrogen in the astronomical epsilon notation: $\varepsilon_E = N(E)/N(H_{Tot})$ for Equation (18) yields

$$K_{Gr} = \frac{a_{Gr}}{P_{Tot}} \times \frac{1 + \varepsilon_{He}}{\varepsilon'_C}$$

(19)

When metallicity is increased, the solar abundance of an element relative to hydrogen is increased by the factor *m*, where $\log(m) = [M/H]$. As metallicity is the relative abundance of all elements heavier than He, the factor m is applied to all elements heavier than He. For a non-solar metallicity, Equation (19) is

$$K_{Gr} = \frac{a_{Gr}}{P_{Tot}} \times \frac{1 + \varepsilon_{He}}{m\varepsilon'_C}$$

(20)

Taking the log of Equation (20) and rearranging

$$\log(K_{Gr}) = \log(a_{Gr}) + \log(1+\varepsilon_{He}) - \log(m\varepsilon'_C) - \log(P_{Tot})$$

(21)

Where graphite is stable, $a_{Gr} = 1$, so Equation (21) becomes

$$\log(K_{Gr}) = \log(1+\varepsilon_{He}) - \log(\varepsilon_C^{'}) - \log(P_{Tot}) - [M/H]$$

(22)

Replacing the equilibrium constant with a linear fit, $\log(K) = A + B/T$,

$$A + \frac{B}{T} = \log(1+\varepsilon_{He}) - \log(\varepsilon_C^{'}) - \log(P_{Tot}) - [M/H]$$

(23)

Rearranging,

$$\frac{1}{T} = \frac{(\log(1+\varepsilon_{He}) - \log(\varepsilon_C^{'}) - A))}{B} - \frac{1}{B}\log(P_{Tot}) - \frac{1}{B}[M/H]$$

(24)

Equation (24) is the general equation for the initial condensation temperature of graphite from monatomic carbon. Using Equation (24), elemental abundances, and $\log(K_i)$ constants A and B, the *analytical* solution for R1 at C/O = 1.2 is:

$$\frac{10^4}{T} = 3.2512 - 0.2671 \log(P_{Tot}) - 0.2671 [M/H]$$

(25)

This relation explains the metallicity and total pressure dependence of the condensation temperature for R1. The coefficient on the metallicity term is identical to the pressure term in the analytical solution, In R1 a similar change in metallicity or total pressure gives a similar change in condensation temperature.

Equation (25) demonstrates that the dependence of initial condensation temperature on total pressure and metallicity can be derived. We provide fits to our full equilibrium calculations in Table 2. Because the chemistry of all elements is coupled, the fits should be used instead of the simplified analytical solutions.

The fit from multivariate regression analysis of $T_{cond}$ C(Gr) from R1 (abundant C(g)) at C/O = 1.2 is:

$$\frac{10^4}{T} = 3.7438 - 0.2267 \log(P_{Tot}) - 0.2604 [M/H]$$

(26), Table 2

In the fit, the coefficients on metallicity and total pressure are similar. A change of the same magnitude in metallicity or total pressure gives a similar change in condensation temperature for this region.

In R2, the second most abundant C-bearing gas species changes from C(g) to $C_2H_2$(g) with increasing pressure and most hydrogen is in monatomic H(g), so graphite forms via the net thermochemical reaction

$$C_2H_2(g) = 2C(Gr) + 2H(g)$$

The $K_{Gr}$ for this reaction is

$$K_{Gr} = \frac{P_H^2 \times a_{Gr}^2}{P_{C_2H_2}} = \frac{P_{Tot}^2 \times X_H^2 \times a_{Gr}^2}{P_{Tot} \times X_{C_2H_2}}$$

(27)

Simplifying,

$$K_{Gr} = \frac{P_{Tot} \times X_H^2 \times a_{Gr}^2}{X_{C_2H_2}}$$

(28)

The $T_{cond}$ C(Gr) decreases with increasing pressure in R2 because the total pressure dependence in the equilibrium constant ($K_{Gr}$) expression is in the numerator, whereas in R1, the pressure term is in the denominator. While an analytical solution can be derived using the same procedure as Equation (25), the simplifications used above do not yield a good approximation because the reaction 2H(g) = $H_2$(g) is proceeding and altering all mole fractions (see Equations (11) and (12)). The empirical fit for this reaction is presented in Table 2:

$$\frac{10^4}{T} = 8.2822 + 0.3704 \log(P_{Tot}) - 1.5004 \, [M/H]$$

(29), Table 2

In R3, the most abundant hydrogen gas species is diatomic $H_2(g)$ and graphite forms via the net thermochemical reaction:

$$C_2H_2(g) = 2C(Gr) + H_2(g)$$

Here, $K_{Gr}$ is

$$K_{Gr} = \frac{fH_2 \times a_{Gr}^2}{P_{C_2H_2}} = \frac{P_{Tot} \times X_{H_2} \times a_{Gr}^2}{P_{Tot} \times X_{C_2H_2}} = \frac{X_{H_2} \times a_{Gr}^2}{X_{C_2H_2}}$$

(30)

In R3, the pressure dependencies cancel out because the number of moles of gas on each side of the reaction are equal. In this region, the flat curvature of the $T_{cond}$ C(Gr) curve can intersect the carbide condensation temperature curves and change the condensation sequences (Figures 3, 5, 7). In the $C_2H_2$ and $H_2$ regions, the graphite initial condensation temperature is almost solely dependent on metallicity. The analytical solution for R3 is:

$$\frac{1}{T} = \frac{(-\log(\varepsilon_C') - A)}{B} - \frac{1}{B}[M/H]$$

(31)

$$\frac{10^4}{T} = 5.7335 - 0.8685 [M/H]$$

(32)

The fit given in Table 2 has a pressure dependence due to the coupled chemistry of all elements, however this dependence is hundreds of times smaller than the metallicity dependence.

$$\frac{10^4}{T} = 5.8140 + 0.0013 \log(P_{Tot}) - 0.8541 [M/H]$$

(33), Table 2

In R3a, R3b, and R3c, graphite forms via the same net thermochemical reaction $C_2H_2(g) = 2C(Gr) + H_2(g)$. The $T_{cond}$ C(Gr) is pressure independent in R3a and R3b, however there is a pressure dependence in R3c, see §3.2.2.

The equal abundance boundary between C and $C_2H_2$ shifts to higher pressures at higher metallicities, changing the total pressure where the condensation sequence matches that observed in presolar grains. For the lowest metallicities (-1.0, -1.5, and -2.0), $C_2H_2$ is the second most abundant C-bearing gas over nearly the entire pressure range shown. For intermediate metallicities (-0.5, 0.0, and 0.5), the C to $C_2H_2$ change is the only speciation change at temperatures where graphite condensation begins. As shown in figure 5, the carbon speciation becomes increasingly complex as metallicity increases, causing the broad peak of the $T_{cond}$ C(Gr) curves for [M/H] = +1.0, +1.5, and +2.0. Beginning with [M/H] = +1.0, the $C_3$ region forms and grows with increasing metallicity at temperatures where graphite condensation occurs, though only at a narrow temperature and total pressure range (Figure 6A-C).

Unlike C and $C_2H_2$, the effect of $C_3$ as the second most abundant C-bearing gas is not obviously seen in the $T_{cond}$ C(Gr) curves. The similar slope of $T_{cond}$ C(Gr) curves where C or $C_3$ is the second most abundant C-bearing gas is best explained through the analytical solution for condensation temperature. Where $C_3(g)$ is more abundant than $C(g)$, graphite forms via the net thermochemical reaction:

$$C_3(g) = 3C(Gr)$$

The analytical solution for initial condensation temperature where $C_3$ is the second most abundant C-bearing gas is:

$$\frac{10^4}{T} = 3.5651 - 0.2349\log(P_{Tot}) - 0.2349[M/H]$$

(34)

These coefficients are comparable to those in Equations (25) and (26) so that no inflection point is seen where carbon speciation changes from C to $C_3$ as the second most abundant C-bearing gas. As the regions are not distinguishable and $C_3$ is only a significant gas at very high metallicity, we give the C and $C_3$ region as one fit in Table 2, Equation (26).

3.2.2 Carbide Condensation

As described in §2, where carbides condense after graphite, the partial pressures of C-bearing gases are fixed and the C/O ratio in the gas no longer matches the bulk C/O ratio (Lodders & Fegley 1997a). Carbide initial condensation temperature dependence on metallicity, total pressure, and graphite stability is easily seen in the fits in Table 2. In the equation for SiC condensation after graphite, total pressure and metallicity have essentially equal coefficients, thus similar changes in metallicity and total pressures have nearly equal effects on initial condensation temperature:

$$\frac{10^4}{T} = 4.6039 - 0.3672 \log(P_{Tot}) - 0.3614 [M/H]$$

(35), Table 2

Where SiC condenses before graphite, metallicity has a larger effect on initial condensation temperature than total pressure when either is changed by the same factor:

$$\frac{10^4}{T} = 4.8570 - 0.2892 \log(P_{Tot}) - 0.4679 [M/H]$$

(36), Table 2

The same effect is observed in TiC initial condensation temperatures. Where graphite is stable:

$$\frac{10^4}{T} = 4.3092 - 0.2985 \log(P_{Tot}) - 0.2895 [M/H]$$

(37), Table 2

Where graphite is not yet stable:

$$\frac{10^4}{T} = 4.5513 - 0.2517 \log(P_{Tot}) - 0.3771 [M/H]$$

(38), Table 2

An important consideration of carbide condensation is the effect carbide stability can have on graphite condensation temperatures. As seen in §3.2.1, there is a pressure dependence in $T_{cond}$ C(Gr) in R3c, where SiC condenses before graphite. When carbon is sequestered in SiC, the $T_{cond}$ C(Gr) decreases with increasing pressure because removal of C into SiC also lowers the partial pressure of $C_2H_2$. This effect only occurs at low metallicities (-1.0, -1.5, and -2.0) and high total pressures where SiC condenses before graphite (Figure 3). Titanium is not in high enough abundance to lower the partial pressure of C-bearing gases significantly when TiC condenses before graphite.

3.2.3 Condensation Temperature Fits

Analytical solutions described in §3.2.2 are for illustrative purposes, but empirical linear regression fits to the results of the CONDOR code calculations should always be used to predict condensation temperatures. The fits are provided in Table 2 for -10 ≤ log(P/bar) ≤ -4, -2 ≤ [M/H] ≤ +2, and C/O = 1.2 provide the condensation temperatures with the maximum uncertainty noted if the graphite curves are more than Δ log(P/bar) = ±0.1 from a gas speciation change. Generally, the further from a speciation boundary, the smaller the uncertainty will be. Care should be taken to note the C/O, total pressure, and metallicity ranges where each equation applies.

Table 2. Initial condensation temperature fits from linear regression for temperature in K[4,5].

| TiC-SiC-Graphite (R3c) | $\frac{10^4}{T} = A + B \log(P_{Tot}) - C [M/H]$ | | | Maximum uncertainty (K) |
|---|---|---|---|---|
| | A | B | C | |
| TiC (before graphite) | 4.5513 | -0.2517 | -0.3771 | ±5 |
| SiC (before graphite) | 4.8570 | -0.2892 | -0.4679 | ±10 |
| Graphite ($C_2H_2$ and $H_2$) | 6.1345 | 0.0933 | -0.9771 | ±15 |
| TiC-Graphite-SiC (R3b) | | | | |
| TiC (before graphite) | 4.5513 | -0.2517 | -0.3771 | ±5 |
| SiC (after graphite) | 4.6039 | -0.3672 | -0.3614 | ±10 |
| Graphite ($C_2H_2$ and $H_2$) | 5.8140 | 0.0013 | -0.8541 | ±1 |
| Graphite-TiC-SiC (R1, R2, R3a) | | | | |
| TiC (after graphite) | 4.3092 | -0.2985 | -0.2895 | ±5 |
| SiC (after graphite) | 4.6039 | -0.3672 | -0.3614 | ±10 |
| Graphite (R1, C or $C_3$ and H) | 3.7438 | -0.2267 | -0.2604 | ±15 |
| Graphite (R2, $C_2H_2$ and H) | 8.2822 | 0.3704 | -1.5004 | ±15 |
| Graphite (R3a, $C_2H_2$ and $H_2$) | 5.9919 | 0.0346 | -0.9171 | ±10 |

---

[4] for $-2.0 \leq [M/H] \leq +2.0$ and $-10 \leq \log(P/bar) \leq -4$ at C/O = 1.2

[5] Indicated in parentheses is the region of Figures 4 and 5 that the equations correspond to.

3.2.4 Condensation Temperatures at Different C/O Ratios

Figure 7 compares condensation temperatures at three different C/O ratios. With increasing C/O, the condensation temperatures increase. Carbide condensation temperatures vary ~20 K between the lowest and highest C/O ratio for the same pressures, so they are slightly sensitive to C/O. Carbide condensation is essentially C/O ratio independent if carbides condense after graphite because the activity of graphite is fixed at unity, see §3.2.2. Graphite condensation temperatures are sensitive to C/O because an increase in carbon abundance is comparable to an increase in metallicity for only carbon and because C-bearing gas speciation is sensitive to C/O. The total pressure where the change from C(g) to $C_2H_2$(g) as the second most abundant C-bearing gas occurs increases with C/O, similar to the effect of increasing metallicity described in §3.1. The $T_{cond}$ C(Gr) peak is broad at the highest metallicities due to the complexity of the carbon gas chemistry, as shown in Figure 6C for C/O = 1.2. The TiC-C(Gr)-SiC region shifts to higher pressures and lower metallicities as C/O increases. At C/O = 3.0, the condensation sequence between -1.0 ≤ [M/H] ≤ +1.0 is always C(Gr)-TiC-SiC for -10 ≤ log(P/bar) ≤ -4, while at C/O = 1.1, all three of the possible condensation sequences occur.

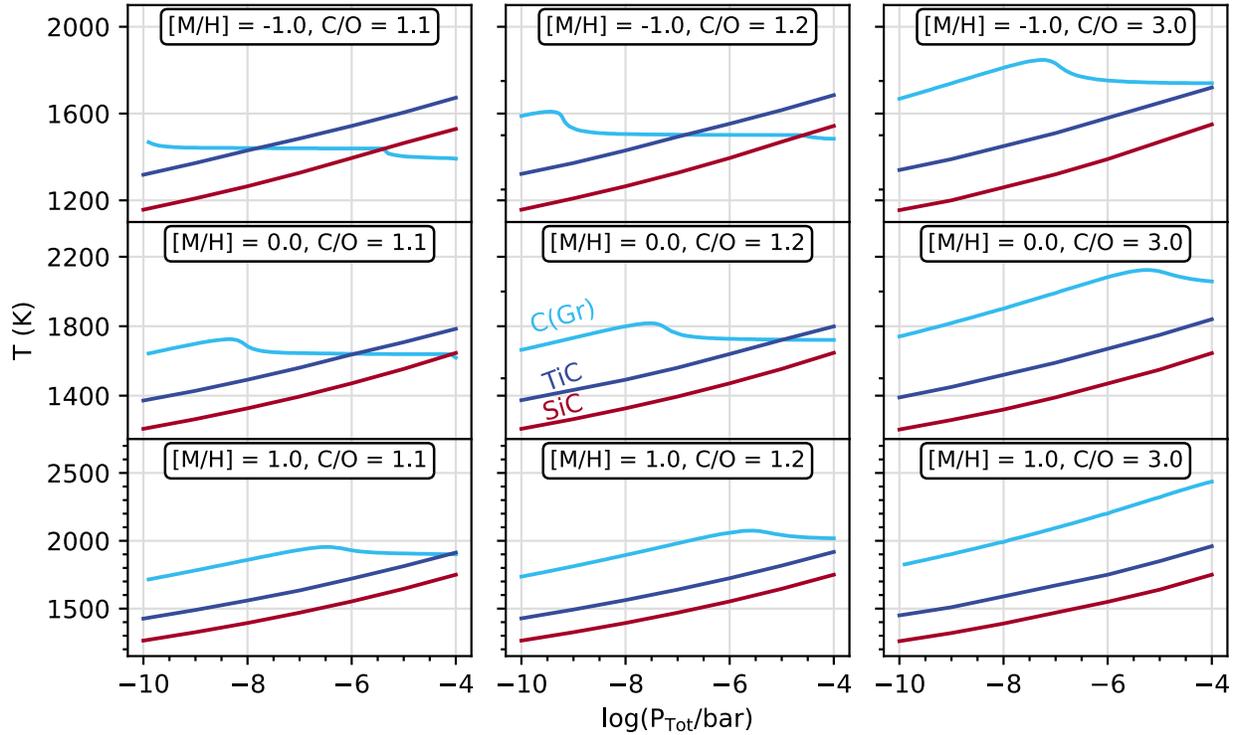

Figure 7. Initial condensation temperatures at C/O ratios of 1.1, 1.2, and 3.0 for the metallicities of -1.0, 0.0, and 1.0.

3.3 Using Effective Carbon to Estimate the Sequence

As shown in §3.2 and §3.2.1, the TiC-C(Gr)-SiC sequence only occurs where $C_2H_2$ is the second most abundant C-bearing gas. The following relationship between available carbon ($\varepsilon'C$), which is the ratio of available carbon to hydrogen (e.g., Equations (18) and (19)), and C-bearing gas speciation can be used to predict the total pressures where regions R1, R2, and R3 occur. Estimating these regions is useful to approximate the sequence for other C/O ratios.

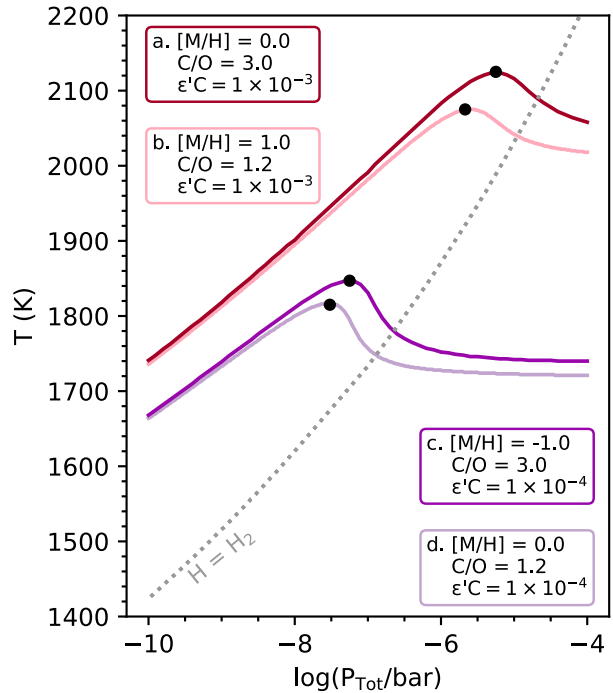

*Figure 8. Selected graphite initial condensation temperature curves for different conditions as indicated in the boxes. Black dots indicate the lowest total pressure where $C_2H_2$ is the major carbon-bearing gas after CO (defined as R3 in Figure 5A-C). For a constant $\varepsilon'C$, the total pressures where R3 occurs is approximately the same (within $\Delta log(P/bar) = \pm 0.5$ or less) for any C/O and [M/H].*

A larger $\varepsilon'C$ promotes the formation of less hydrogenated species, whereas a smaller $\varepsilon'C$ promotes more saturated hydrocarbon formation. The total pressure where $C_2H_2$ becomes the second most abundant gas after CO at temperatures where carbon condensation commences (R3) is closely related to $\varepsilon'C$, with a small dependence on C/O and [M/H] (Figure 8). The four lines in Figure 8 have different metallicities and C/O ratios, but where $\varepsilon'C$ is the same (8a and 8b, 8c and 8d), the $C_2H_2$ region occurs at approximately the same pressures. In all cases except 8b, the black dot also indicates where $X(C) = X(C_2H_2)$. For 8b, the metallicity is supersolar, so there is a small range of pressures in between the C and $C_2H_2$ regions where $C_3$ and HCN are the second most abundant C-bearing gas (Figure 6C). For 8b, the $C_3$ and HCN regions occur between $-6.7 \leq log(P/bar) \leq -6.5$.

For $1\times10^{-5} < \varepsilon'C < 1\times10^{-3}$ and the pressure range $-10 \leq \log(P/bar) \leq -4$, the empirical relationship from linear regression for the lowest total pressure where R3 occurs, with an uncertainty of $\Delta\log(P/bar) = \pm 0.1$, is:

$$\log(P_{R3}) = 2.38827 - 0.62099\,[M/H] + 2.52033\,\log(\varepsilon'C) - 0.60591\,\log(C/O)$$

(39)

For $\varepsilon'C < 1\times10^{-5}$, all carbon near the graphite condensation temperatures is in $C_2H_2$ over the $-4 \leq \log(P/bar) \leq -10$ range. For $1\times10^{-3} < \varepsilon'C < 1\times10^{-2}$, $C_2H_2$ is only favored as the second most abundant C-bearing gas as the pressure approaches $\log(P/bar) = -4$, and other C-bearing species can be the second most abundant C-bearing gas in-between monatomic C and $C_2H_2$, such as HCN and $C_3$ (Figure 8b; Figure 6C). If $10^{-2} < \varepsilon'C$, $C_2H_2$ is never the favored C-bearing gas within the pressure range considered here and the sequence is always C(Gr)-TiC-SiC.

*4. Discussion*

*4.1 Major Gases Relevant to Grain Formation*

Gas chemistry in the circumstellar envelopes is metallicity dependent and determines the condensation sequence of carbonaceous condensates. At all C/O ratios and metallicities investigated, depending on total pressure, the second most abundant C-bearing gas (after CO) is nearly always monatomic C or $C_2H_2$. $C_3$ can only be the second most abundant C-bearing gas when metallicities are much higher than the range where the condensation sequence is TiC-C(Gr)-SiC, and only for a very small P-T range (Figure 6C).

Previous condensation calculations for carbon stars were done at solar metallicity, and we compare general trends of our results to other studies. However, the nominal cases from these studies are difficult to summarize because of the different parameters used, such as C/O ratios, thermochemical data, solar abundances, and sometimes P-T profiles. The TiC-C(Gr)-SiC sequence for solar metallicity was investigated using equilibrium calculations for C/O > 1 by Sharp & Wasserburg (1995). They also found that graphite can condense before a significant amount of hydrogen is in $H_2$, but carbides cannot. Our results describe how the association of hydrogen plays a crucial role in determining the condensation sequence. The lowest pressure for the sequence TiC-C(Gr)-SiC is approximately the pressure where the $X(H) = X(H_2)$ line intersects with the graphite condensation temperature curve (Figure 6A-C).

Sharp & Wasserburg (1995) also reported the importance of $C_2H_2$ as a source gas for graphite condensation. However, unlike in our results, they found that $C_2H$ was the second most abundant gas at pressures below $10^{-7}$ bar. Similarly, Chigai et al. (1999) performed equilibrium calculations and reported $C_2H_2$ and $C_2H$ as the most important C-bearing gases for graphite condensation, with a $C_2H$ dominated region at low pressures for C/O > 1 at solar metallicity. Sharp & Wasserburg (1995) and Chigai et al. (1999) largely used the JANAF tables (Chase et al. 1985). The difference between our results and those of Sharp & Wasserburg (1995) and Chigai et al. (1999) originates with the thermochemical data for $C_2H$ gas, as noted in §2. $C_2H$ is less stable than assumed in Chigai et al. (1999) and Sharp & Wasserburg (1995). Chigai et al. (1999) note that $C_2H$ is expected to be a photodissociation product of $C_2H_2$ and, observationally, is found in the outer circumstellar envelope (Keady &

Hinkle 1988; Agúndez et al. 2017; Unnikrishnan et al. 2024). The lower stability of $C_2H$ gas given by more recent thermochemical data resolves the puzzle of why $C_2H$ gas is not observed in the inner CSE. Our predictions of C and $C_2H_2$ as more probable precursors to graphite condensation are observationally supported.

The nominal carbon star calculations in Agúndez et al. (2020) and Salpeter (1977) agree with our finding that monatomic C and $C_2H_2$ are the most important gases for graphite condensation occurring in the inner circumstellar envelope. Both Sharp & Wasserburg (1995) and Salpeter (1977) also identified $C_3$ as an important C-bearing gas for condensation in low pressure regions where C is the second most abundant C-bearing gas. We do not find that $C_3$ becomes more important than C or $C_2H_2$ for graphite condensation at or below solar metallicity, but we do find that $C_3$ is a significant gas in low pressure regions at high metallicity.

The gas and condensation chemistry presented in here and in §3.1 define permissible ranges in pressure, temperature, and metallicity to explain the graphite grains containing refractory titanium carbide inclusions studied in the laboratory. These computations can be used to constrain circumstellar envelope conditions. For a given C/O and P-T profile, the relative abundance of C, $C_3$, $C_2H_2$ gases at distances where carbonaceous dust formation commences can be used as a metallicity indicator (see Figure 2 and §3). A higher abundance of $C_2H_2$ instead of C or $C_3$ indicates a lower metallicity than the same C/O and P-T profile with a lower abundance of $C_2H_2$. Carbonaceous dust obscures the inner circumstellar environment in the near-infrared range, complicating observational studies of gases around carbon stars (Stewart et al. 2016; Wittkowski et al. 2017). However, if the dust formation radius for a particular star is known, these computations are known to infer the P-T conditions and major gases in the inner circumstellar environment. Alternatively, the gas chemistry near the radius where carbonaceous condensation commences could be used to constrain grain formation radii if pressure and temperature profiles were known.

Gas chemistry may play a role in determining carbonaceous grain structure, as suggested by Croat et al. (2005) and Messenger et al. (1998) in reference to presolar graphite grains containing polycyclic aromatic hydrocarbons (PAHs) in their cores. These grain cores were first found by Bernatowicz et al. (1996), encased by well-ordered graphitic rims. Figure 6 shows the second most abundant C-bearing gas at a given P-T point, note that complete mass balance of total carbon between gas species and condensates is considered in all computations. As $C_2H_2$ is a frequently invoked precursor for PAH formation (see §4.3), the graphite grain cores may have formed in the $C_2H_2$-rich inner region of the circumstellar envelope. The cores could have then traveled to lower pressure regions further from the star, where monatomic C gas condensed on the cores, forming the graphitic rim.

*4.2 Metallicity of Presolar Grain Parent Stars*

Our results indicate that carbonaceous presolar grains with refractory carbide inclusions likely formed from carbon stars of subsolar to solar metallicity and high to intermediate pressures where the second most abundant C-bearing gas is either monatomic C or $C_2H_2$. At high metallicities, carbon and hydrogen are mostly in monatomic species C (or at especially high metallicity, $C_3$) and H at temperatures and pressures where graphite begins condensing, making the sequence C(Gr)-TiC-SiC. As metallicity decreases, $C_2H_2$ and $H_2$ become the major species where graphite begins condensing. Only where $C_2H_2$ is the second most abundant C-bearing gas can the graphite condensation temperature curve intersect with the carbide curves, changing the sequence to TiC-C(Gr)-SiC and, eventually at the highest pressures and lowest metallicities, TiC-SiC-C(Gr).

As shown in §3.2.1, §3.2.2, and from the fit parameters in Table 2, the dependencies of condensation temperature on metallicity and total pressure are not always the same. Previously, the effect of metallicity on condensation temperatures was assumed to be analogous to the effect of total pressure (Leisenring et al. 2008; Croat et al. 2010). However, the change in heavy element abundances relative to H and He can cause fundamental changes in gas chemistry that are not necessarily analogous to total pressure effects.

As described in §1, metallicities for carbonaceous presolar grain carbon star sources have been predicted to have been as low as [M/H] = -0.85 or as high as [M/H] = +0.3 (Gail et al. 2009; Lugaro et al. 2018,2020). Our independent results using the TiC-C(Gr)-SiC condensation sequence indicate that graphite grains with refractory carbide inclusions are more plausibly explained by origin from stars of subsolar metallicity, agreeing with the conclusions of Gail et al. (2009) and Cristallo et al. (2020). There is only a narrow pressure range where the TiC-C(Gr)-SiC sequence can occur at high metallicities. The supersolar metallicities inferred by Lewis et al. (2013) are between 0 ≤ [M/H] ≤ +0.2. Lugaro et al. (2018, 2020) predicted metallicities as high as [M/H] = +0.3. Our results indicate that at [M/H] = 0, the TiC-C(Gr)-SiC sequence occurs at -5 ≤ log(P/bar) ≤ -4. At [M/H] = +0.3, the TiC-C(Gr)-SiC sequence occurs at -4.6 ≤ log(P/bar) ≤ -4. Such high total pressures directly correspond with high temperatures. Using polarimetric, photometric, spectroscopic, and interferometric observational studies, total pressures in the circumstellar envelope of AGB stars are usually inferred to be log(P/bar) = -4 at the highest (Wong et al. 2016; Khouri et al. 2016,2018,2024; Vlemmings et al. 2017,2019; Ohnaka et al. 2017; Adam & Ohnaka 2019). Time-dependent carbon star atmosphere modeling results are in good agreement with mass loss rates, wind velocities, spectroscopic, and photometric observations, and indicate that these high total pressures correspond to high temperatures (T > 2000 K) (Jorgensen et al. 1992; Nowotny et al. 2011,2013; Eriksson et al. 2014). In our results, we find no supersolar metallicity where both total pressures are favorable to the TiC-C(Gr)-SiC sequence and temperatures are low enough for TiC condensation to start before graphite.

*4.3 Nucleation Timescales*

Carbon dust nucleation has been extensively investigated by equilibrium models (e.g., Salpeter 1974,1977), which predict graphite as a major condensate from carbon stars. Equilibrium studies also constrain the equilibrium gas phase composition, as Cherchneff & Barker (1992) computed for polycyclic aromatic hydrocarbons abundances in carbon stars. Equilibrium studies were also used in combination with kinetic considerations to determine grain formation rates, outflow rates, and grain sizes (e.g., Hoyle & Wickramasinghe 1962; Friedemann & Schmidt 1967; Donn et al. 1968; Kamijo 1969; Bernatowicz et al. 2005), all of which conclude that carbonaceous dust can form in the circumstellar envelopes of carbon stars, providing grains to the interstellar medium. However, Bernatowicz et al. (2005) could not match presolar grain sizes observed in the laboratory when assuming mass loss occurred in spherically symmetrical outflows, instead suggesting high pressure asymmetries in stellar outflows could be the site of nucleation. Dense "clumps" of dust have been observed in carbon-rich stellar outflows, supporting this theory (e.g., Stewart et al. 2016; Ohnaka et al. 2016). Chemical kinetic pathways toward carbonaceous dust have been described by Frenklach & Feigelson (1989), Keller (1987), Tielens (1990), Tielens (2022), and Cherchneff et al. (1992). Kinetic departures from equilibrium in circumstellar environments have been discussed in detail by Nuth & Donn (1981) and Donn & Nuth (1985), who concluded that classical nucleation theory is not sufficient to describe the nucleation of circumstellar grains. Here, we estimate the timescales for initial nucleation at initial condensation temperature for TiC, SiC, and graphite as an upper limit on grain formation times.

Carbon dust formation in stellar ejecta has long been suspected to form through the same chemical reaction pathways as terrestrial soot formation: polycyclic aromatic hydrocarbons (PAHs) are the intermediate between acetylene and soot (Salpeter 1974; Keller 1987; Frenklach & Feigelson 1989; Tielens 1990; Cherchneff & Barker 1992; Tielens 2022). This conclusion is supported by evidence of PAHs in the cores of graphite grains (Bernatowicz et al. 1996; Messenger et al. 1998).

Several pathways to PAHs have been identified for temperatures relevant to circumstellar envelopes, typically beginning with $C_2H_2$, as acetylene is generally the most abundant hydrocarbon in terrestrial combustion and very abundant in carbon star circumstellar envelope conditions (Pentsak et al. 2024). Johansson et al. (2018) suggested a radical-driven hydrocarbon clustering mechanism wherein resonance stabilized radical species covalently bond to $C_2H_2$ or $C_2H_3$, producing a clustered product. The lower limit estimate for reaction collision efficiency ($\alpha$) for these low-barrier soot formation reactions involving PAHs and radicals is 0.01 (Zhou et al. 2022) and the upper limit is barrierless reactions with an $\alpha$ of 1.

The mechanisms for SiC and TiC formation in circumstellar envelopes are less understood. The stability and cluster-growth of TiC and SiC and their potential observational features were studied by Gobrecht et al. (2017) and Gámez-Valenzuela et al. (2021). For SiC, the lower limit of reaction collision efficiency is estimated to be 0.1 (Yasuda & Kozasa 2012). For TiC, the measured reaction collision efficiency in microgravity experiments is 0.29 (Kimura et al. 2023).

When nucleation timescales are less than the expansion and pulsation variability timescales[6], equilibrium can be applied (Kamijo 1969; Gail & Sedlymayr 1988; Lodders & Fegley 1997b, 2025). The residence time for material in the dust formation zone is on the order of years (see review by Gail and Sedlmayr 2013), which is to be compared to typical variability cycles of around a year for Mira variables. We use the equation derived in Gail & Sedlmayr (1988) for the approximate chemical timescale, $\tau_{ch}$, for dust formation and growth:

$$\tau_{ch}=1/(A_n\, v\, n_0 \alpha)$$

(40)

Where $A_n$ is the reactive area in $cm^2$, $v$ is velocity in cm/s, $n_0$ is particle density in molecules/$cm^3$, and $\alpha$ is reaction collision efficiency.

---

[6]The pulsation of the luminosity variability cycle (on the order of years) should not be confused with the thermal pulses (on the order of 10,000 years or greater).

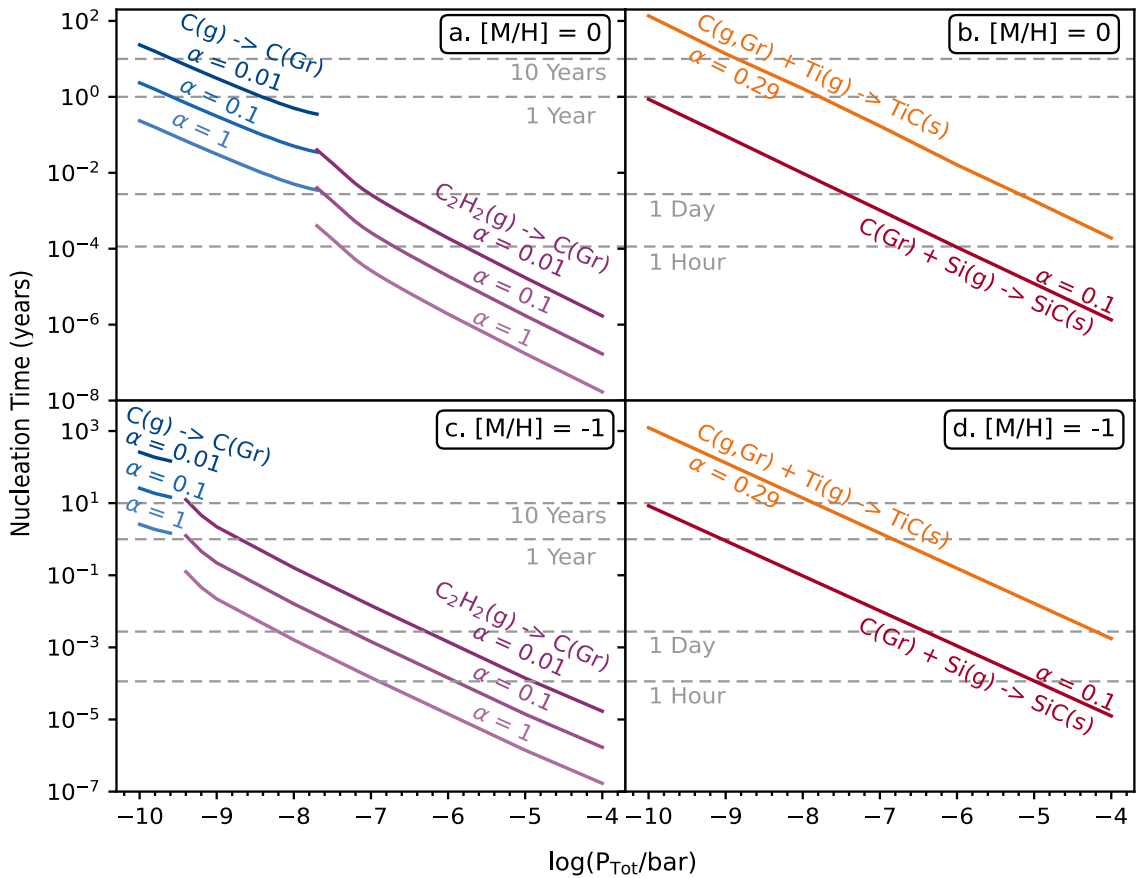

*Figure 9. Initial nucleation times at initial condensation temperatures of graphite, TiC, and SiC at [M/H] = 0 (a and b) and -1 (c and d) for C/O = 1.2. Where initial nucleation times are $10^1$ years or less, equilibrium calculations can be used to approximate the condensation sequence.*

Figure 9 shows timescale approximations using Equation (40) for initial nucleation of graphite (Figures 9a and 9c), and carbides (Figures 9b and 9d). The velocity *v* is on the order of $10^4$ cm/s (Gail & Sedlmayr 1988), $n_0$ comes from the mole fraction at the initial condensation temperature at [M/H] = 0.0 and -1.0 from this study. The major carbon source gas is either monatomic C or $C_2H_2$, as described in Figure 6B. For graphite condensation, the number density of monatomic C or $C_2H_2$ is used. For $C_2H_2$, using the molecular diameter from Hirschfelder et al. (1964), the reactive area $A_1$ is $5.6\times10^{-15}$ cm². For monatomic C, using the atomic diameter from Slater (1964), the reactive area $A_1$ is $6.2 \times10^{-16}$ cm². For TiC and SiC, the equilibrium Ti(g) or Si(g) number densities at the carbide initial condensation temperatures are used. The timescales reported here for TiC and SiC are calculated with the monatomic C diameter for reactive area, as it is the smallest reactive area, to give an upper limit on timescale. Carbide nucleation times are potentially much faster.

In the case of graphite condensation at [M/H] = 0, the initial nucleation time is under one year for all but the very lowest limit of estimated reaction collision efficiency at very low pressures (Figure 9a). In the case of SiC, initial nucleation always occurs in less than one year (Figure 9b). For TiC, the initial nucleation timescale is greater than one year where log(P/bar) < -7.8 (Figure 9b). However, for TiC to condense before graphite at solar metallicity, log(P/bar) must be above -5 (Figure 5B), where TiC initial nucleation timescales are ~1 day or less, thus the long timescale at low pressures is not a concern.

We also examine the subsolar case [M/H] = -1, where timescales are longer by approximately an order of magnitude. Graphite initial nucleation from monatomic C does not happen within one year at any reaction collision efficiency, which is only a concern at log(P/bar) < -9.45 (Figure 9c). Where log(P/bar) > -9.45, graphite nucleates from $C_2H_2$ within 10 years or less. SiC always nucleates on the order of years or less. For TiC to condense before graphite at [M/H] = -1, total pressures need to exceed log(P/bar) > -7 (Figure 5A), where timescales are 1 year or less. Thus, we conclude that thermochemical equilibrium applies to our results because in all cases but the lowest total pressures for TiC and the lowest efficiency/low pressure case for graphite, the chemical kinetic timescale for dust nucleation is less than the residence time, which is on the order of years (Gail & Sedlmayr 2013).

*5. Conclusions*

Presolar graphite grains with TiC and other refractory trace element carbide inclusions imply that the condensation sequence is TiC, then graphite, then SiC. Our results suggest that these TiC-bearing presolar graphite grains formed in carbon stars of subsolar to solar metallicities at pressure ranges around -8 ≤ log(P/bar) ≤ -5.

The highest metallicity where this sequence occurs for C/O = 1.2 is at [M/H] = +0.5, but only at the highest total pressure (log(P/bar) = -4) The temperatures corresponding to such high total pressures in the circumstellar envelopes are likely too high for stable condensates. At and below solar metallicity, the TiC-C(Gr)-SiC sequence is favored at high to intermediate pressures within a circumstellar envelope, where temperatures would be more favorable for refractory carbide and graphite condensation (e.g., Jorgensen et al. 1992; Lodders & Fegley 1997b).

Increasing the C/O ratio increases the pressures and decreases the metallicities where the TiC-C(Gr)-SiC sequence can occur. At C/O = 1.2, this sequence is already limited to intermediate and high pressures and metallicities at or below solar. Thus, higher bulk C/O ratios increase the required total pressure and metallicity conditions for the TiC-C(Gr)-SiC sequence higher than the total pressure and metallicity ranges that are plausible for carbon stars from which the carbide-bearing graphite grains originated (Jorgensen et al. 1992; Gail et al. 2009).

We have shown that stellar metallicity is an important parameter for condensation in circumstellar envelopes of carbon stars. The effects of metallicity on condensation temperatures are unique from total pressure and C/O ratio effects. Stellar metallicity should be considered carefully before applying solar metallicity calculations to other stars.

Acknowledgements

This work was supported by the McDonnell Center for the Space Sciences and NSF AST-2108172. We thank B. Fegley for thorough and insightful comments, as well as B. Jolliff and H. Chou for computing resources. We also thank the anonymous reviewer for detailed comments which have improved this manuscript.40